\documentclass[a4paper,11pt]{article}
\pdfoutput=1 

\usepackage{float} 
\usepackage{graphicx}
\usepackage{dcolumn}
\usepackage{bm}
\usepackage[T1]{fontenc}
\usepackage{ulem}
\usepackage{lineno}

\newcommand{\NUCLEUS}{\textsc{Nucleus} }
\newcommand{\FIFRELIN}{\textsc{FIFRELIN} }

\newcommand{\CRESST}{\textsc{Cresst} }
\newcommand{\EDELWEISS}{\textsc{Edelweiss} }
\newcommand{\GEANT}{\textsc{GEANT4} }
\newcommand{\CEvNS}{\textsc{CE$\nu$NS} }
\newcommand{\nuc}[2] {$^{\text{#1}}$#2}

\usepackage{jinstpub} 

\title{Calibration of nuclear recoils at the 100 eV scale using neutron capture}

\author[a]{L.~Thulliez}
\author[a,1]{D.~Lhuillier\note{Corresponding author.}}
\author[b]{F.~Cappella}
\author[b]{N.~Casali}
\author[c,d]{R.~Cerulli}
\author[a]{A.~Chalil}
\author[e]{A.~Chebboubi}
\author[a]{E.~Dumonteil}
\author[f]{A.~Erhart}
\author[g]{A.~Giuliani}
\author[a]{F.~Gunsing}
\author[h]{E.~Jericha}
\author[f]{M.~Kaznacheeva}
\author[f]{A.~Kinast}
\author[f]{A.~Langenkämper}
\author[a,f]{T.~Lasserre}
\author[a]{A.~Letourneau}
\author[e]{O.~Litaize}
\author[g]{P.~de Marcillac}
\author[g]{S.~Marnieros}
\author[a]{T.~Materna}
\author[a]{B.~Mauri}
\author[a]{E.~Mazzucato}
\author[a]{C.~Nones}
\author[f]{T.~Ortmann}
\author[d,i]{L.~Pattavina}
\author[g]{D.V.~Poda}
\author[a]{R.~Rogly}
\author[f]{N.~Schermer}
\author[e]{O.~Serot}
\author[a]{G.~Soum}
\author[j]{L.~Stodolsky} 
\author[f]{R.~Strauss}
\author[b,k]{M.~Vignati}
\author[a]{M.~Vivier}
\author[f]{V.~Wagner}
\author[f]{A.~Wex}

\affiliation[a]{IRFU, CEA, Universit\'e Paris-Saclay, 91191 Gif-sur-Yvette, France}
\affiliation[b]{INFN—Sezione di Roma, Piazzale Aldo Moro 2, I-00185, Roma, Italy}
\affiliation[c]{INFN, Sezione di Roma “Tor Vergata”, I-00133 Roma, Italy}
\affiliation[d]{Dipartimento di Fisica, Università di Roma “Tor Vergata”, I-00133 Roma, Italy}
\affiliation[e]{CEA, DES, IRESNE, DER, Cadarache F-13108 Saint-Paul-Lez-Durance, France}
\affiliation[f]{Physik-Department, Technische Universität München, D-85748 Garching, Germany}
\affiliation[g]{Universit\'e Paris-Saclay, CNRS/IN2P3, IJCLab, 91405 Orsay, France}
\affiliation[h]{TU Wien, Atominstitut, 1020 Wien, Austria}
\affiliation[i]{INFN, Laboratori Nazionali del Gran Sasso, 67100 Assergi (AQ), Italy}
\affiliation[j]{Max-Planck-Insitut für Physik, D-80805 München, Germany}
\affiliation[k]{Sapienza Università di Roma, Dipartimento di Fisica, I-00185 Roma, Italy}

\date{\today}

\emailAdd{david.lhuillier@cea.fr}

\abstract{The development of low-threshold detectors for the study of coherent elastic neutrino-nucleus scattering and for the search for light dark matter necessitates methods of low-energy calibration. We suggest this can be provided by the nuclear recoils resulting from the $\gamma$ emission following thermal neutron capture. In particular, several MeV-scale single-$\gamma$ transitions induce well-defined nuclear recoil peaks in the 100 eV range. Using the FIFRELIN code, complete schemes of $\gamma$-cascades for various isotopes can be predicted with high accuracy to determine the continuous background of nuclear recoils below the calibration peaks. We present a comprehensive experimental concept for the calibration of CaWO$_4$ and Ge cryogenic detectors at a research reactor. For CaWO$_4$ the simulations show that two nuclear recoil peaks at 112.5 eV and 160.3 eV should be visible above background simply in the spectrum of the cryogenic detector. Then we discuss how the additional tagging for the associated $\gamma$ increases the sensitivity of the method and extends its application to a wider energy range and to Ge cryogenic detectors.}


\arxivnumber{2011.13803} 

\begin{document}
\maketitle
\flushbottom

\section{Introduction}
\label{sec:intro}
Recent advances in the development and operation of ultra-sensitive particle detectors have opened a wide window of opportunities in neutrino and dark matter physics. This community effort involves studying nuclear recoils at very low energies in the order of 10 eV. These could be produced either by low-mass dark matter (DM) candidates~\cite{Kaplan:1991ah,Feng:2008ya,Essig:2012yx,PhysRevD.85.063503} or by  coherent elastic neutrino-nucleus scattering (CE$\nu$NS) \cite{Freedman:1977xn,Drukier:1983gj}. A thorough understanding of the detector response at these energy scales is therefore of the utmost importance. In particular, for these processes the interaction rates increase strongly at low energies. Thus,  the community  needs  a procedure for the accurate reconstruction of the energy for low-energy nuclear recoils.  Advanced devices such as scintillating detectors, ionization detectors and noble-liquid-based time projection chambers suffer from  imprecision in the energy determination of nuclear recoils at low energy, due to the poorly understood energy quenching~\cite{Lindhard:1961zz,Ovchinnikov2004,Birks_1951}. In particular, the emerging technology of low-threshold cryogenic detectors lacks  a suitable calibration procedure at sub-keV energies, both for electron and nuclear recoil events.  The usual method involves an extrapolation of the detector energy response over several orders of magnitude~\cite{Manalaysay:2009yq,PhysRevLett.112.171303}. Few data were obtained in the sub-keV energy range with Ge crystals in the ionization channel \cite{PhysRevA.11.1347,Collar:2021fcl}. They show an unclear situation on the evolution of the quenching factors, reflecting the difficulty of these measurements and the need for complementary approaches.

\begin{figure}[htbp]
    \begin{center}
	\includegraphics[width=0.9\linewidth]{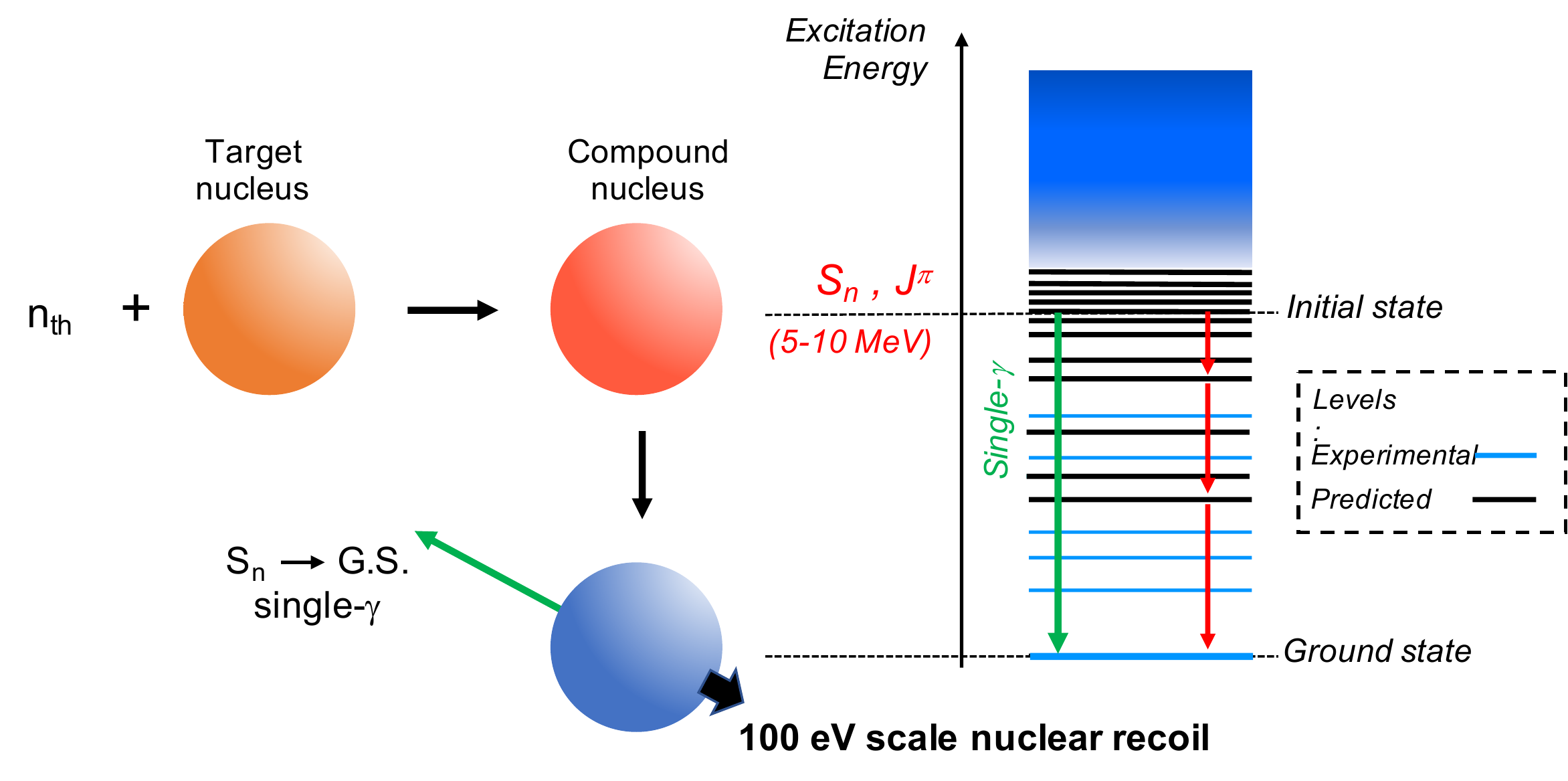}
	\end{center}
	\caption{\label{fig:FIFRELIN} Illustration of the process of radiative neutron capture. The sought-for signal is the nuclear recoil associated with a single-$\gamma$ transition from the $S_n$ level to ground state. The two-body kinematics determine a unique nuclear recoil energy. The complete distribution of single and multi-$\gamma$ decays from all isotopes is predicted by the FIFRELIN simulations, combining experimental level schemes and predictions from  level density models.}
\end{figure}

In this paper, we propose the direct measurement of nuclear recoils induced by the capture of thermal neutrons for an accurate calibration in the $100$\,eV range. The  kinetic energy of the captured neutron is negligible ($\approx$ 25 meV) and the capture creates a compound nucleus in an excited state close to the neutron separation energy $S_n$, of several MeV. This nucleus then de-excitates to the ground state via $\gamma$ and conversion electron emission. Although there is thus typically a multi-$\gamma$ cascade, there is often a single-$\gamma$ transition with a substantial probability directly to the ground state (figure~\ref{fig:FIFRELIN}). The resulting nuclear recoil ($E_{\gamma}^{2}/2M$) has an energy in the order of 100 eV for medium and heavy mass nuclei. The high energy photon easily escapes cm-size detectors. This escape has two implications  for our method. One is that the only energy remaining in the detector is that of the nuclear recoil. Secondly, adding  a $\gamma$ detector allows the event to be ``tagged'' with this photon, so that the additional coincidence can  strongly suppress backgrounds. Below we shall present examples with and without this tagging.

Among the various experimental techniques where this calibration method could be applied, cryogenic particle detectors have a high potential. They are ideal for the validation of this procedure~\cite{leo1991}, having a very low energy threshold, a few eV, and a small active volume, a few cm$^3$ ~\cite{Strauss:2017cuu, Abdelhameed:2019hmk,Armengaud:2019kfj}. They  currently lead the field of low-mass direct DM investigations ~\cite{Abdelhameed:2019hmk,Angloher:2017sxg,Arnaud:2020svb,Armengaud:2019kfj}, and are also proposed as primary candidates for the study of CE$\nu$NS~\cite{Strauss:2017cuu,Billard:2016giu,Angloher:2019flc}.

\begin{figure}[htbp]
    \begin{center}
	\includegraphics[width=0.8\linewidth]{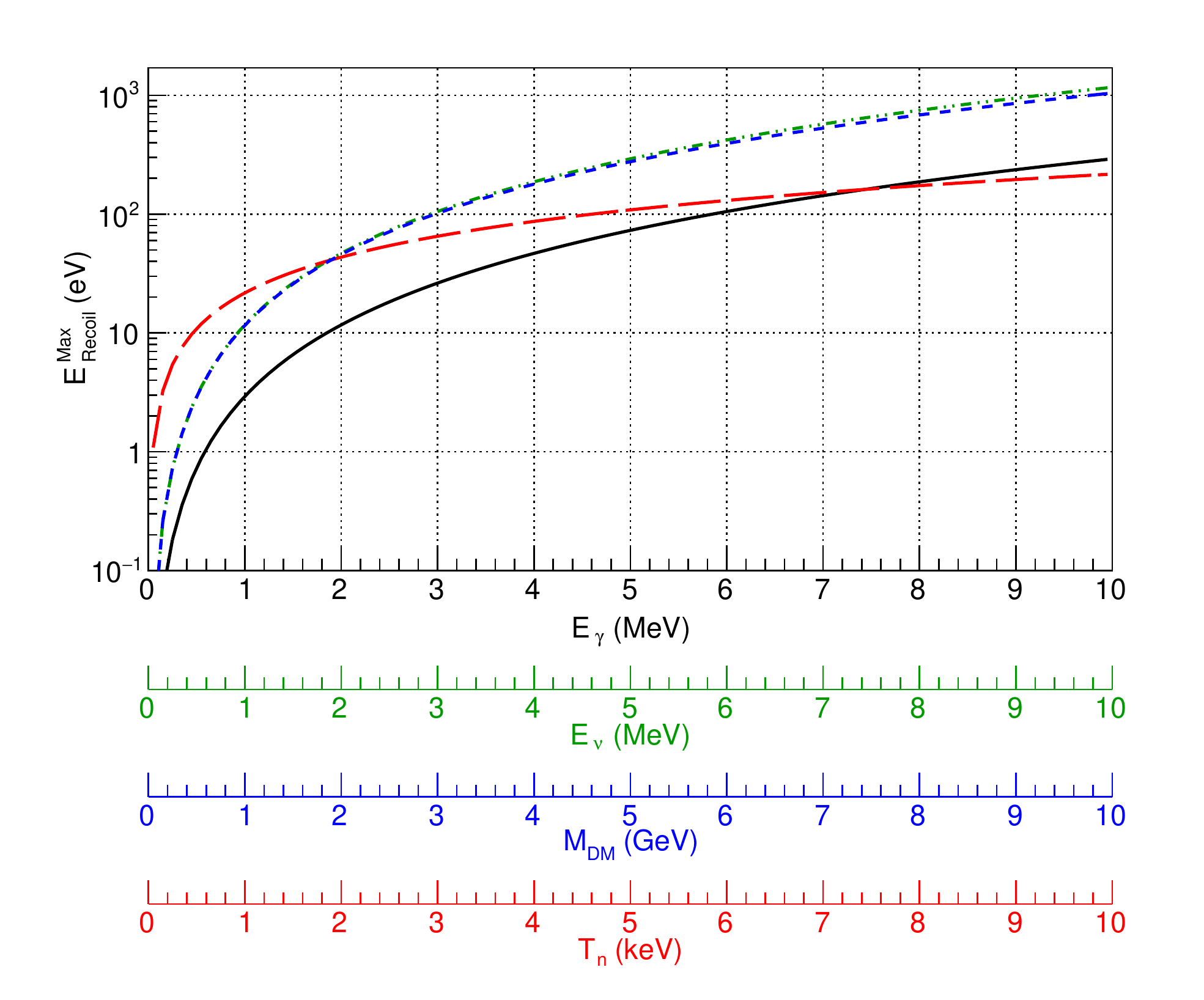}
	\end{center}
	\caption{\label{fig:Recoils_W} Nuclear recoils  induced by the emission of a $\gamma$-ray of energy $E_\gamma$ (solid black curve), compared with the maximum recoils in CE$\nu$NS with  incident neutrino energy $E_\nu$ (dashed-dotted green) and with  recoils from elastic scattering of a dark matter particles of mass $M_{DM}$. The dark matter particles are taken to have a speed of 300 km/s, most probable value of the speed distribution expected in the galaxy (short-dashed blue curve). It turns out that these choices of DM speed and horizontal axes make the DM and neutrino curves virtually identical. Also shown are the maximum recoils induced by elastic scattering of neutrons with kinetic energy $T_n$ (long-dashed red curve). The mass of the nucleus has been taken for that of tungsten, to a good approximation, all recoil energies scale with the inverse mass of the target nucleus.}
\end{figure}

Figure \ref{fig:Recoils_W} compares the nuclear recoil energies expected for light dark matter,  CE$\nu$NS and elastic neutron scattering with that for  single-$\gamma$ transitions. One notes that the recoil energies for the different processes cover the same range. The mass of the nucleus has been taken as that of tungsten, the recoil energy scales as the inverse mass  of the nucleus.

\section{Principle of the method}
\label{sec:principle}

\subsection{Properties for target nuclei}
\label{subsec:key-param}
A suitable target nucleus for the technique presented here should have a high natural abundance ($Y_{ab}$) and a high thermal neutron capture cross-section ($\sigma_{n,\gamma}$). The compound nucleus should have a sizeable branching ratio for the single-$\gamma$ transitions ($I_{\gamma}^{s}$). We thus define a figure of merit (FoM) as the product of these three quantities.
\begin{equation}\label{eq:FoM}
  FoM=\sigma_{n,\gamma}\times Y_{ab}\times I_{\gamma}^{s}.
\end{equation}
Table \ref{tab:nuc_data} shows the nuclear data for W and Ge, which are of interest for a number of ongoing and future experiments~\cite{Strauss:2017cuu,Abdelhameed:2019hmk,Billard:2016giu,Agnolet:2016zir,Pattavina:2020cqc} using CaWO$_4$, PbWO$_4$ or Ge detectors. However, it is worth  mentioning that this calibration procedure may be applied to other problems, provided  the  nuclei involved  have suitable properties.
\begin{table}
\caption{\label{tab:nuc_data} Nuclear data for W and Ge isotopes (see text for notations). \nuc{180}{W} is not considered here because of its low natural abundance of 0.12\%. The neutron capture cross section are for thermal neutrons.}
\begin{center}
\begin{tabular}{|r|rr|rrr|r|}
\hline \hline
\multicolumn{3}{|c|}{Target nucleus ($A$)} & \multicolumn{3}{c|}{Compound nucleus ($A$+1)} &  \tabularnewline
\hline
Isotope & $Y_{ab}$ \cite{Isotopiccompositionsoftheelements2013IUPACTechnicalReport} & $\sigma_{n,\gamma}$ \cite{ENDF/B-VIII.0} & $S_n$ \cite{EGAFpubli} & $I_{\gamma}^{s}$ \cite{EGAFpubli, Hurst2014} & Recoil & FoM \tabularnewline
   & (\%) & (barn) & (keV) & (\%) & (eV) & (a.u.) \tabularnewline
\hline
 \nuc{182}{W} & 26.50 & 20.32 & 6191 &  13.94  & 112.5 & 7506 \tabularnewline
 \nuc{183}{W} & 14.31 & 9.87 & 7411 &  5.83 & 160.3 & 823  \tabularnewline
 \nuc{184}{W} & 30.64 & 1.63 & 5754 &  1.48 & 96.1 & 74  \tabularnewline
 \nuc{186}{W} & 28.43 & 37.89 & 5467 &  0.26  & 85.8 & 280  \tabularnewline
\hline
 \nuc{70}{Ge} & 20.53 & 3.05 & 7416 &  1.95  & 416.2 & 122  \tabularnewline 
 \nuc{72}{Ge} & 27.45 & 0.89 & 6783 &  0.0 & 338.7 & 0  \tabularnewline
 \nuc{73}{Ge} & 7.76 & 14.70 & 10196 &  0.0 & 754.9 & 0  \tabularnewline
\nuc{74}{Ge}  & 36.52 & 0.52 & 6506 &  2.83 & 303.2 & 54  \tabularnewline
\nuc{76}{Ge}  & 7.74 & 0.15 &  6073 &  0.0 & 257.3 & 0  \tabularnewline
\hline \hline 
\end{tabular}
\end{center}
\end{table}

\subsection{Prediction of the de-excitation cascades by the FIFRELIN code}
\label{subsec:fifrelin-pred}
Although our principal interest is the high energy single $\gamma$'s from the excited energy $S_n$ in the compound nucleus, we must also consider the full cascade following neutron capture, since this could be an important source of background and is also relevant for rate/deadtime considerations.

To predict the spectrum of the associated  nuclear recoils we employ the \FIFRELIN code developed at CEA (France). This code has been validated in different physics cases (see e.g. \cite{Litaize2015,Almazan2019}) and has also shown good agreement with independent calculations performed with the DICEBOX code \cite{Becvar1998}. To build the level scheme of a nucleus, \FIFRELIN first includes all available experimental data (energy levels, spins, parities, branching ratios, half-lives when available) from the RIPL-3 \cite{RIPL3}, ENSDF \cite{ENSDF} or EGAF \cite{EGAFpubli} databases. Since a complete level scheme is not experimentally accessible, it is completed by sampling levels (see figure \ref{fig:FIFRELIN}) in the nuclear level density Composite Gilbert-Cameron model \cite{RIPL3}. Briefly this model is a combination of a Fermi gas model at high energy around the neutron separation energy, and the constant-temperature model at lower energy. Within the Fermi gas model, the level density is simply related to the effective excitation energy through $\rho_{FG} \propto \exp(\sqrt{aU})/U^{5/4}$, where $a$ stands for the so-called level density parameter based on the Ignatyuk prescription \cite{Ignatyuk1975} and $U=E-\Delta$ stands for an effective excitation energy accounting for pairing energy $\Delta$ related to the odd-even effects in nuclei. At low energy, the level density within the constant temperature model reads $\rho_{CT}=\exp((E-E_0)/T)/T$, where $E_0$, the energy shift, and $T$, the nuclear temperature, are free parameters determined by fitting experimental data such as the cumulated number of levels. The level spin is sampled following the work of Bethe \cite{Bethe_1936,Bethe_1937} and summarized in the RIPL-3 documentation: $P(J)=((2J+1)/2\sigma^2)\ \exp(-(J+1/2)^2/2\sigma^2)$, where $\sigma^2$ is the spin cut-off parameter that can be constant or energy-dependent. Finally, both positive and negative parities are assumed to be equally distributed. The gamma transition probabilities between levels are computed from the Enhanced Generalized Lorentzian (EGLO) radiative strength function model \cite{RIPL3} for $E1$ transitions and the Standard Lorentzian (SLO) model for other multipolarities. The parameters involved in these models have been measured for some nuclei, as for the gadolinium main isotopes, otherwise they are deduced from global systematic parameterizations, as simple functions of mass number for example. The conversion electrons are accounted for through tabulated internal conversion coefficients using the BrIcc code \cite{Kibedi2008}. A transition between two levels is sampled among all possible transitions, fully characterizing the emitted particles ($E,J,\pi$). This is repeated until the ground state is reached. Thus by construction all $\gamma$-cascades have the correct total energy. More details on the algorithm can be found in \cite{Regnier2016}. 

In order to perform a calculation, FIFRELIN needs as input the initial nuclear state ($E$, $J$, $\pi$) of the compound nucleus ($A$,$Z$) formed by neutron capture. If the reaction selection rules for spin and parity lead at thermal energy to the superposition of two possible initial states $J_1^\pi$ and $J_2^\pi$, two simulations are performed weighted by the relative spin contributions to the thermal neutron capture cross-section. 
Finally, FIFRELIN performs multiple nuclear de-excitations on several sampled level schemes and associated partial transitions to account for the systematic uncertainty of the model resulting from our lack of knowledge on the entire nucleus level scheme. Such a set of calculations is called “Nuclear Realization”.

\begin{figure}[htbp]
    \begin{center}
	\includegraphics[width=0.8\linewidth]{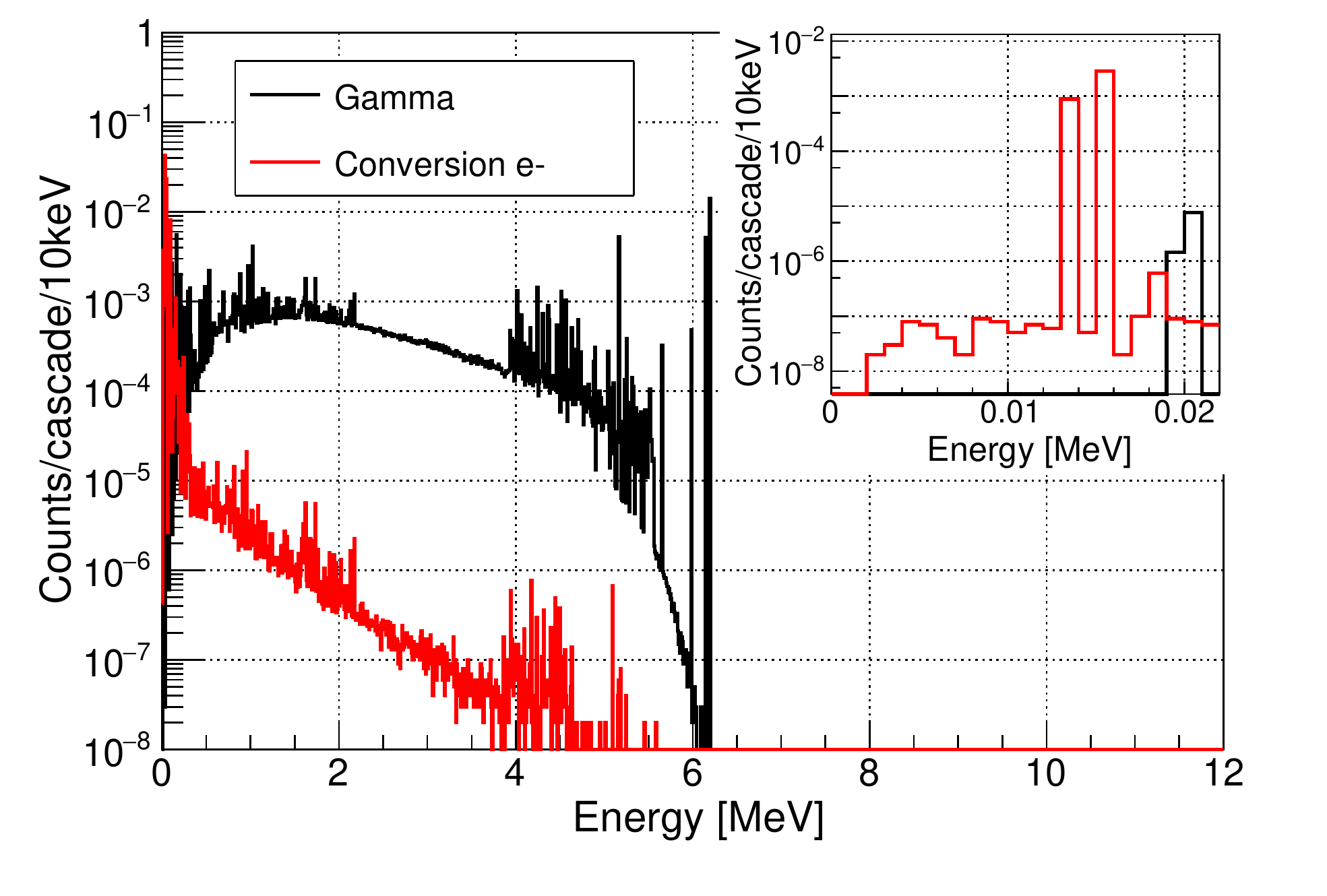}
	\end{center}
	\caption{\label{fig:183W_Spectra} Gamma (black) and conversion electron (red) spectra from the de-excitation of an individual $^{183}$W atom following a thermal neutron capture on $^{182}$W, as predicted by the FIFRELIN code with experimental levels taken from the RIPL-3 and EGAF databases. Inset: zoom of the same curves at very low energy showing no conversion electrons below 2 keV and no $\gamma$-rays below ~20 keV.}
\end{figure}

As an example, the predicted gamma and conversion electron spectra emitted by an individual atom of $^{183}$W (corresponding to the highest FoM in table~\ref{tab:nuc_data}) are shown in figure~\ref{fig:183W_Spectra}. The $\gamma$-line of highest energy, at 6.191 MeV, corresponds to single-$\gamma$ transitions from the excitation energy at $S_n$ to the ground-state, inducing mono-energetic (112.5 eV) recoils of the de-exciting $^{183}$W nucleus. This line is intense and well separated from the rest of the $\gamma$-spectrum, favoring a strong calibration signal. A closer look to the spectrum reveals a neighboring line at 6.144 MeV, which has to come with a 47~keV secondary $\gamma$-ray to connect to the ground state. However, the resulting difference with respect to the previous recoil line is at the \%-level and most of the time the 47~keV photon will lead to an energy deposit in the bolometer above the RoI. Thus the impact of this secondary $\gamma$-line is negligible and the mean recoil energy remains accurately predicted.

The spectrum is smooth at intermediate energies (2 to 4 MeV) because of the huge amount of transitions arising from the multiple combination of spins and parities of initial and final states. This is not the case at low energies where the main contribution comes from the lower part of the nuclear level scheme with discrete levels clearly identified. Similarly primary transitions have been identified at high energy through the EGAF database and are accounted for in the simulation. These transitions are fully characterized by their known initial state ($S_n$) and final state (the lower part of level scheme has been measured).

\subsection{Expected calibration peaks and prompt internal background}
\label{subsec:calib-peak}
Similar simulations of $\gamma$-cascades can be run for all isotopes in a bolometer. Thus in the following FIFRELIN predictions are coupled with detailed \GEANT simulations for the $\gamma/e^{-}$  propagation and neutron captures in the crystal. As a first  example we consider the cryogenic CaWO$_4$ detector of the \NUCLEUS experiment. The crystal is a 5\,mm length cube (0.76 g mass), and has achieved energy thresholds of $\sim20$\,eV along with an energy resolution of 3.4\,eV \cite{Strauss:2017cam,Angloher:2017sxg}. We safely take  5\,eV in the following simulations. A beam of thermal neutrons is sent perpendicular to one face of the cube. 

The thermal neutron beam and the transport of all emitted photons and electrons from neutron-capture vertices are simulated with \GEANT version 10.04.p02~\cite{GEANT4_2016}. The neutron\_HP package based on ENDF/B-VII nuclear data evaluation \cite{ENDF/B-VII.0} and the NCrystal library \cite{Cai2020} are used to describe the neutron interactions and EMZ physics list for the electromagnetic processes. We find that 10.5\% of the  neutrons are captured,  quasi-uniformly in the detector volume. Due to the low capture cross-sections of calcium and oxygen, 97.6\% of the captures are on the tungsten isotopes, listed in table~\ref{tab:nuc_data}. Among all captures in the crystal 5\% lead to the emission of a single-$\gamma$.  The consequent mono-energetic nuclear recoil provides a clean calibration signal, since it is only rarely accompanied by other simultaneous energy deposits: in fact, the $\gamma$-ray has only a 5\% probability to interact 
in a \NUCLEUS CaWO$_4$ crystal and, as illustrated in figure \ref{fig:183W_Spectra}, has a negligible probability to convert to an electron. Thus the mono-energetic nuclear recoils remain the main source of energy deposit for these events. 

However, most of the nuclear decays following neutron capture involve multiple, lower energy, $\gamma$’s ($\gamma$-cascades). These have a short attenuation length, a high conversion coefficient, and therefore a high probability of interacting in the cm-dimension crystal. In the following, we call “internal background” such energy deposits inherent to the neutron capture process. One might naively expect this background to spoil the signal from the  single-$\gamma$ transitions we wish to use for calibration. But in fact the interactions in the crystal  actually help in rejecting a large fraction of such events. The key point, illustrated in the inset of figure \ref{fig:183W_Spectra}, is that the energies are always large ($\ge$2 keV), so that the deposited energy in a cascade is very rarely  in the range $20<E<200$~eV. We define this as our region of interest (RoI),  encompassing the single-$\gamma$ recoils expected for W isotopes as in table~\ref{tab:nuc_data}. The only exceptions to this argument, as revealed by the simulations, occur at the very low rate of 4.1$\times$10$^{-4}$ events per neutron capture. These are from electron events very close to the crystal surface depositing a small amount of their energy before escaping the crystal, or very forward Compton scattering where the $\gamma$-ray leaves a small energy before escaping. Most of the multi-$\gamma$ cascades lead to large (several keV to MeV) energy deposits from Compton and conversion electrons, moving the majority of the internal backgrounds to far above the RoI. This feature improves the signal-to-background ratio around the calibration peak but the price for this is a higher rate of large signals: considering all capture events, the ratio of number of events in the RoI with respect to the total number of events is 0.13. This induces more dead-time in the slow, low temperature, detector and so constrains the acceptable neutron beam intensity.

\begin{figure}[htbp]
    \begin{center}
	\includegraphics[width=0.8\linewidth]{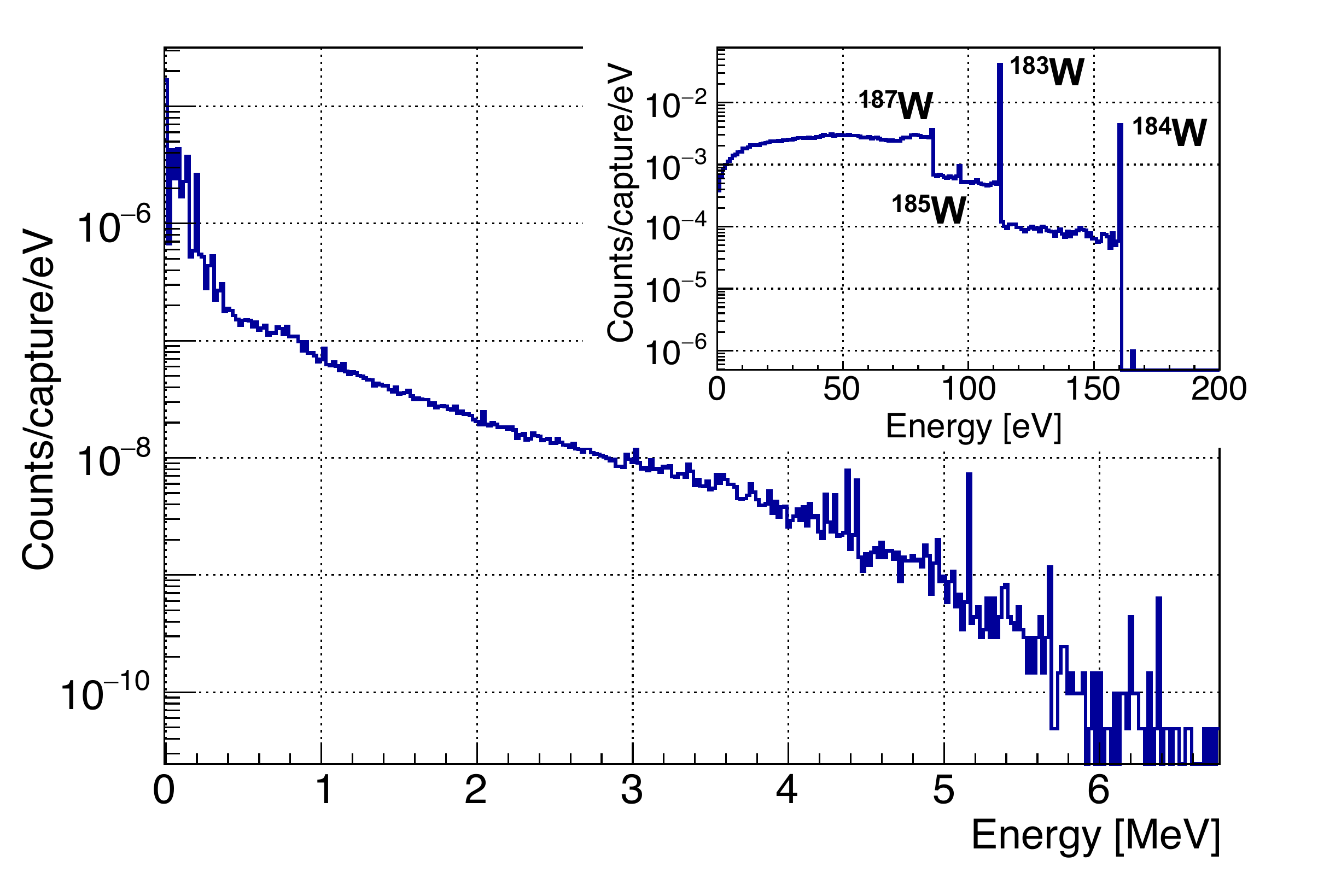}
	\end{center}
	\caption{\label{fig:Edep_Bolo_W} Distribution of energy deposits in a 5\,mm length cubic crystal of CaWO$_4$ based on a simulation with 10$^7$ incident thermal neutrons. Inset: zoom to the RoI showing the expected calibration peaks. No resolution effects  are included.}
\end{figure}

Finally, the RoI is almost exclusively populated by pure nuclear recoils, induced on the  one hand by single-$\gamma$ transitions, which  provide our calibration peaks (5\% of the neutron captures in CaWO$_4$) and on the other hand, by multi-$\gamma$ cascades with no conversion electrons and all $\gamma$-ray(s) escaping the detector without interaction (20\% of the neutron captures). Figure \ref{fig:Edep_Bolo_W} shows the  spectrum of energy deposits obtained from a simulation with 10$^{7}$ thermal neutrons. As expected, high-energy events (MeV scale) are induced by electron recoils from the multi-$\gamma$ cascades. A zoom to the RoI reveals the calibration peaks from the main capture isotopes of tungsten. The nuclear recoils associated with the multi-$\gamma$ cascades without conversion electrons and all $\gamma$'s escaping the crystal, form the continuous distribution to the left of each calibration peak. A clear gap above 160 eV separates these signals from the higher energy depositions. This spectrum demonstrates the potential of the method, with the critical provision  that the detector has a very good energy resolution. 

\subsection{Delayed events}
\label{subsec:delayed}
So far all decays of the compound nucleus have been considered as prompt. We investigate here the possible impact of delayed events coming from the decay of either isomeric states or unstable ground states of the nuclei produced by neutron capture. Indeed these states can have long enough half-lifes with respect to the time response of the bolometer to produce an event separated in time from the prompt event of the neutron capture. The corresponding nuclear data are shown in table \ref{tab:long_lived_states} for tungsten and germanium isotopes. All listed decays emit either electrons or low energy $\gamma$-rays that have a high probability to deposit all their energy in the detector crystal, far above the RoI. Thus their main impact is on the dead-time of the detector. In the worst case, a single neutron capture can induce three separated high-amplitude pulses: one from the multi-$\gamma$ cascade from $S_n$ to an isomeric state, one from the decay of this isomeric state and one from the decay of the unstable ground-state. 
Assuming a typical few days experiment, a detailed calculation accounting for the lifetimes and branching ratios of table \ref{tab:long_lived_states} predicts on average 0.53 delayed events per capture on a tungsten nucleus. This is dominated by the production of \nuc{187}{W} (59\% of the total capture rate occurs on \nuc{186}{W}) having a 24 h half-life. This adds to the prompt events as an extra contribution to the internal background and needs to be taken into account in determining the neutron beam intensity. A similar calculation for the germanium case, discussed in section \ref{subsec:coinc-mode}, has been performed and is found to be 0.35 delayed events per capture on average.
 
\begin{table}[htbp]
\caption{\label{tab:long_lived_states} Nuclear data for the W and Ge isotopes produced by thermal neutron capture, as relevant for the evaluation of delayed signals. The first two columns show the compound nucleus formed by capture and its relative probability, given the natural abundances of table \ref{tab:nuc_data}. The ground state properties are from the ENSDF database \cite{ENSDF}. The isomeric state properties are from FIFRELIN calculation and the RIPL-3 database \cite{RIPL3}. The time evolution of the cascade, from which the isomeric state feeding is computed, uses the Weisskopf estimate based on the single particle approximation and the available data from the RIPL-3 database. EC stands for electron capture.}
\begin{center}
\begin{tabular}{|cc|ccc|ccc|}
\hline \hline
\multicolumn{2}{|c|}{} & \multicolumn{3}{c|}{Ground state} & \multicolumn{3}{c|}{Isomeric state } \tabularnewline
\hline
         Isotope        & Capture    &  Decay   & $T_{1/2}$ & $Q$    & Feeding   & $T_{1/2}$ & Energy \tabularnewline
        ($A$+1)         & [\%] &          &           &  (keV) &  (t>10 ms)[\%]  &   & (keV)     \tabularnewline
\hline
 \nuc{183}{W}  & 30     & -             & stable    & -     & 1     & 5.30 s & 309  \tabularnewline
 \nuc{184}{W}  & 8      & -             & stable    & -     & -     & -     & -     \tabularnewline
 \nuc{185}{W}  & 3      & $\beta^{-}$   & 75.1 d    & 431   & 0.3   & 1.67 min & 197    \tabularnewline
 \nuc{187}{W}  & 59     & $\beta^{-}$   & 24 h      & 1312  & -     & -     & -  \tabularnewline
\hline
 \nuc{71}{Ge}  & 28     & EC            & 11.4 d   & 233   &  9    & 20.4 ms & 198    \tabularnewline 
  \nuc{73}{Ge} & 11     &  -            & stable    &   -   & 45   & 0.499 s   & 67    \tabularnewline 
 \nuc{74}{Ge}  & 52     & -             & stable    & -     & -     & -     & -   \tabularnewline
\nuc{75}{Ge}   & 9      & $\beta^{-}$   & 1.38 h    & 1177  & 24    & 47.7 s & 140    \tabularnewline
\nuc{77}{Ge}   & <1   & $\beta^{-}$   & 11.2 h   & 2703  & -     &   -   &   -     \tabularnewline 
\hline \hline 
\end{tabular}
\end{center}
\end{table}

Some delayed events could be exploited to push the calibration further. In table \ref{tab:long_lived_states} the decay of the ground-state of $^{71}$Ge \textit{via} an electron capture process is of special interest. Here the particle emitted by the nucleus is a mono-energetic neutrino of 233 keV, inducing a negligible recoil of 0.3 eV. However the capture of the L-shell electron is also followed by the emission of a UV-photon, providing an interesting 1.3 keV calibration line from electron recoils (see for instance \cite{Collar:2021fcl,Arnaud:2020svb}).

\subsection{Neutron flux}
\label{subsec:n-flux}
The long decay times of the detector pulses, of the order of 10\,ms, imply stringent constraints on the total counting rate that cryogenic detectors can withstand. Prior to further consideration on the design of the experiment, the distribution of energy deposits resulting from neutron capture in the detector (figure \ref{fig:Edep_Bolo_W}) can be used to determine the optimal intensity of the neutron beam. A dedicated measurement was performed with a $5\times5\times5$ mm$^3$ CaWO$_4$ crystal of the \NUCLEUS experiment. Artificial heater pulses corresponding to the simulated energy spectrum were injected at various rates. An optimal event rate of 10\,Hz  after cuts was found for this gram-scale detector. 
We find that an incident thermal neutron flux of 270\,n/cm$^2$/s (equivalent to 66\,n/s on the crystal) fulfills this specification, with an associated neutron capture rate in the crystal of 6.70 Hz, an internal background rate in the RoI of 1.34 Hz and a rate of 0.32 Hz for the sought-for signal from the single-$\gamma$ transitions (table~\ref{tab:rates}).

\begin{table}[htbp]
\caption{Rates simulated for a 0.76 g CaWO$_4$ crystal (5\,mm length cube) in an incident flux of 270 neutron/cm$^2$/s. When the simulation produces no events, the upper limit is set from the statistical accuracy.}
\label{tab:rates}
\begin{center}
\begin{tabular}{|ll|cc|}
\hline \hline
    & & Total ($s^{-1}$) & RoI ($s^{-1}$) \tabularnewline
    & & $E$ > 20 eV & 20 < $E$ < 200 eV  \tabularnewline
\hline
\multicolumn{2}{|l|}{\bf{Internal Background}} & & \tabularnewline
1)  & Multi-$\gamma$'s & 6.36 & 1.34 \tabularnewline
2)    & Delayed & 3.55 & neglected \tabularnewline
\hline
\multicolumn{2}{|l|}{\bf{External Background}} & & \tabularnewline
3)  & Reactor OFF & 0.16 & 0.08 \tabularnewline
4)  & Reactor ON & & \tabularnewline
    & $\gamma$ ambiance in the hall & 0.05 & $<3\times10^{-3}$ \tabularnewline
    & neutron beam & 0.046 & $<5\times10^{-6}$ \tabularnewline
\hline
\multicolumn{2}{|l|}{\bf{Signal}} & & \tabularnewline
    & Single-$\gamma$'s & 0.34 & 0.32 \tabularnewline
\hline \hline
\end{tabular}
\end{center}
\end{table}

\section{Feasibility Study}
\label{sec:feasibility}
We now consider the practical implementation of this calibration method. Two experimental variants are possible. In the first, the bolometer is sufficient in itself to reveal the sought-for nuclear recoil peak above background. In the second, the emitted $\gamma$(s) are detected in coincidence with the nuclear recoil. Somewhat surprisingly, we find that this first, simpler, variant is possible with CaWO$_4$, thanks to the strong calibration line from  the $^{183}$W isotope (see section \ref{subsec:calib-peak}). That is, the nuclear recoils can be seen, even without $\gamma$-tagging, relying simply on the high energy resolution of the cryogenic detector. This measurement will allow a direct calibration of the \NUCLEUS detectors with nuclear recoils at energies relevant for \CEvNS \cite{Angloher:2019flc}.

\subsection{Experimental setup}
\label{subsec:exp-setup}
\begin{figure}[ht]
    \begin{center}
	\includegraphics[width=0.95\linewidth]{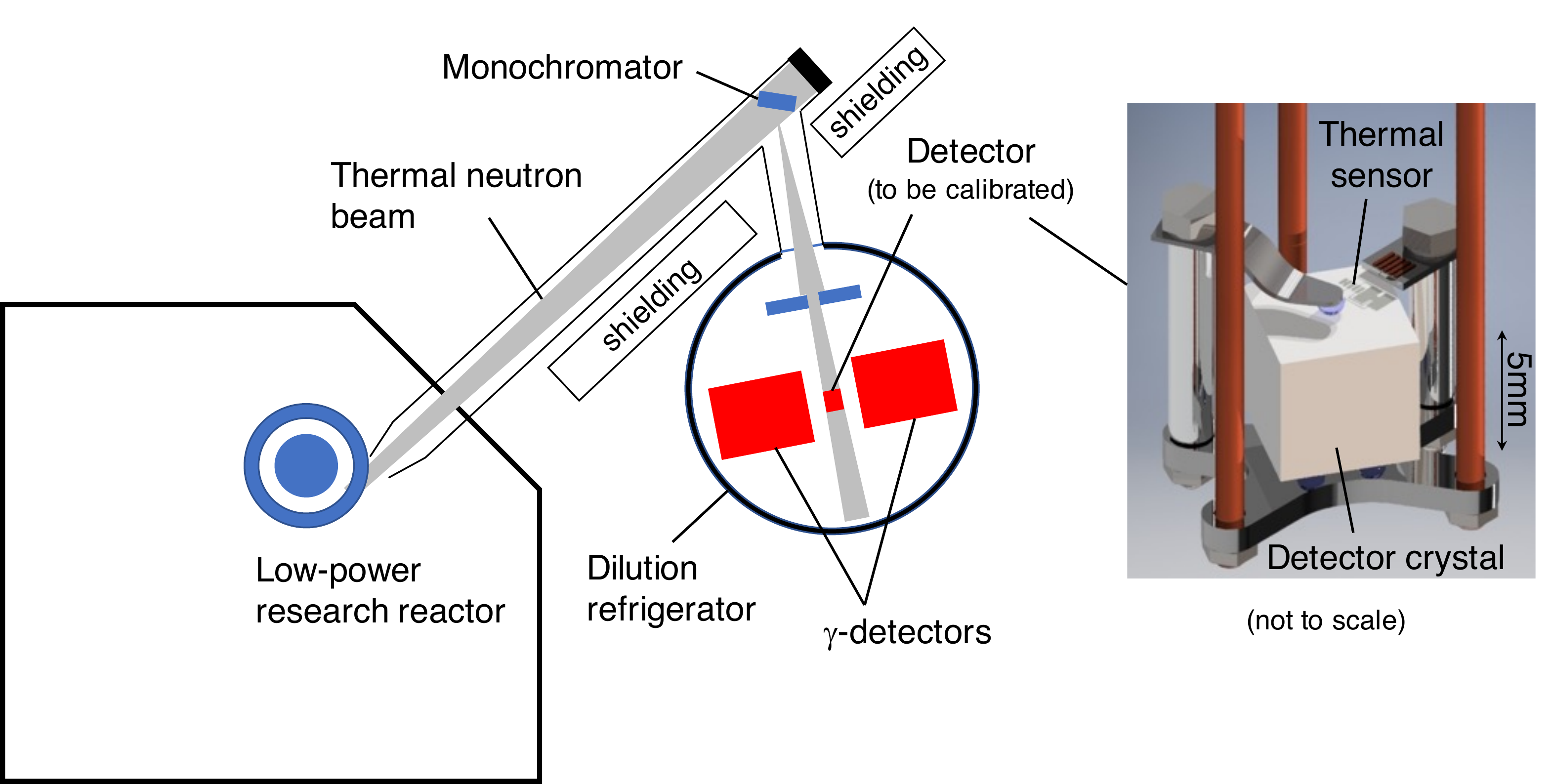}
	\end{center}
	\caption{Illustration of the experimental concept using a low-power TRIGA-Mark II reactor to produce a low intensity, directional and purely thermalized neutron beam. Inset: example detector assembly in the cryostat with thermal sensor and low mass holder to reduce n-induced backgrounds. Sizes of the refrigerator (30 cm inner diameter in reality) and the detector assemblies are magnified for clarity.}
	\label{fig:Exp_Concept}
\end{figure}

A high-quality neutron beam with low background despite the very low required flux would be available at a low-power reactor such as the 250 kW TRIGA-Mark II reactor operating in Vienna \cite{international2016iaea}. We present simulations for a setup at this reactor. Beamlines at this facility  deliver a thermal neutron flux of 10$^{4}$~n/cm$^{2}$/s. Bragg diffraction from a monochromator crystal inserted into the  beamline, in combination with corresponding beam optics, can  provide a collimated neutron beam of the desired flux, with no need for further attenuation, see e.g. \cite{Erhart2012,Jericha2003}. The layout of the neutron beamline and the cryostat proposed here is shown in figure~\ref{fig:Exp_Concept}. The bolometer is in the center of the cryostat. The detection in coincidence of the emitted $\gamma$-rays is studied with two cylindrical detectors placed on both sides of the bolometer (see section \ref{subsec:coinc-mode}).

From this experimental configuration we can discuss the sources of background associated with the reactor site and the detector setup. We label them as “external” background as opposed to the internal background induced by the neutron captures in the bolometer itself.

\subsection{External background}
\label{subsec:ext-bckg}
External backgrounds will include: “Reactor OFF” backgrounds, cosmic rays induced activity and environmental $\gamma$'s, and “Reactor ON” backgrounds, from the neutron beam interacting in the surroundings or ambient activity from the other beam lines in the hall. The “Reactor OFF” background has been measured at the level of $\sim10$\,events/day/eV for a \NUCLEUS-like target detector \cite{Strauss:2017cam,Angloher:2017sxg} under conditions very close to those we are considering, at earth's surface and with no specific shielding around the cryostat. Thus this data, shown as the blue curve in figure \ref{fig:CaWO4_Signal}, is taken as a safe estimate for the present study but it can of course be measured at the real site when the reactor is switched off.

The situation for the “Reactor ON” background is more complicated. First there are the backgrounds induced by operation of the other experiments in the reactor hall. Thermal neutrons from the different beams scatter on the components of the beamlines generating an ambient gas of thermal neutrons and secondary $\gamma$-rays from their capture in surrounding materials. Both components have been  measured at the Vienna site. With the reactor at full power, a flux of thermal neutrons of (0.38$\pm$0.08) n/cm$^2$/s was obtained in the reactor hall using a VacuTec counter tube of 21 mm inner diameter, 100 mm effective length and filled with $^3$He gas at 4 bar. Although this flux is 100 times higher than during “Reactor OFF” periods, its absolute value remains very small with respect to the foreseen beam intensity of 270 n/cm$^2$/s. However, to avoid neutron capture close to the bolometer, the refrigerator could be wrapped with a 5 mm thick borated rubber mat, which proved to efficiently suppress the neutron flux when installed around the He-tube. The ambient $\gamma$-spectrum was measured under the same conditions with an Ortec GMX30-PLUS-S high resolution germanium detector ($\oslash$56.4$\times$61.8 mm$^3$ Ge crystal). This spectrum was unfolded using a simulated response matrix of the Ge crystal and the “true” spectrum  then  injected back into the \GEANT simulation of the experiment. Despite a significant impact of the reactor operation on the $\gamma$ ambiance, the small size of the detector crystal keeps the $\gamma $ background negligible (total rate of 0.05~event/s) with respect to the internal background. No energy deposit at all was recorded in the RoI by the simulation, setting an upper limit of 3$\times$10$^{-3}$ event/s.  

As shown by the red curve in figure \ref{fig:Recoils_W}, a third component of background could be induced by the activity of the reactor, that is the nuclear recoils induced by elastic scattering of neutrons with a few keV kinetic energy. Such fast neutrons could be extracted from the reactor pool by the neighboring beamlines and scattered back toward our experimental setup. However no significant signal of fast neutrons could be measured on site. Using techniques of biased propagation of particles in \GEANT we could show that the direct flux of fast neutrons from the core through the pool and the concrete wall is attenuated by 13 orders of magnitude, a negligible level.

Finally, the last contribution to the “Reactor ON” backgrounds is the interaction of the neutron beam itself with the mechanics of the experiment upstream and downstream of the bolometer. Dedicated \GEANT simulations, including the crystal holder illustrated in figure~\ref{fig:Exp_Concept} and the multi-layer structure of the refrigerator, were performed. A simple configuration of boron-loaded screens collimates the neutron beam upstream of the crystal and stops it downstream. The mm-thickness of copper and aluminum vessels in the refrigerator induces spurious scattering and capture. The bolometer itself scatters  a small fraction of the neutrons, while most of them are absorbed by the boron nuclei in the beam-dump. All secondary $\gamma$-rays generated from neutron captures by nuclei other than in the bolometer are propagated. We find that very few of them (0.046 event/s) deposit energy in the bolometer and none of them in the RoI, leading to an upper limit of 5$\times$10$^{-6}$ count/s in the RoI. The flux of residual fast neutrons in the primary beamline will be highly suppressed by the monochromator. All contributions to signal and background are summarized in table~\ref{tab:rates}.

\subsection{Expected calibration signal in a \NUCLEUS CaWO$_4$ crystal}
\label{subsec:single-mode}
\begin{figure}[ht]
    \begin{center}
	\includegraphics[width=0.8\linewidth]{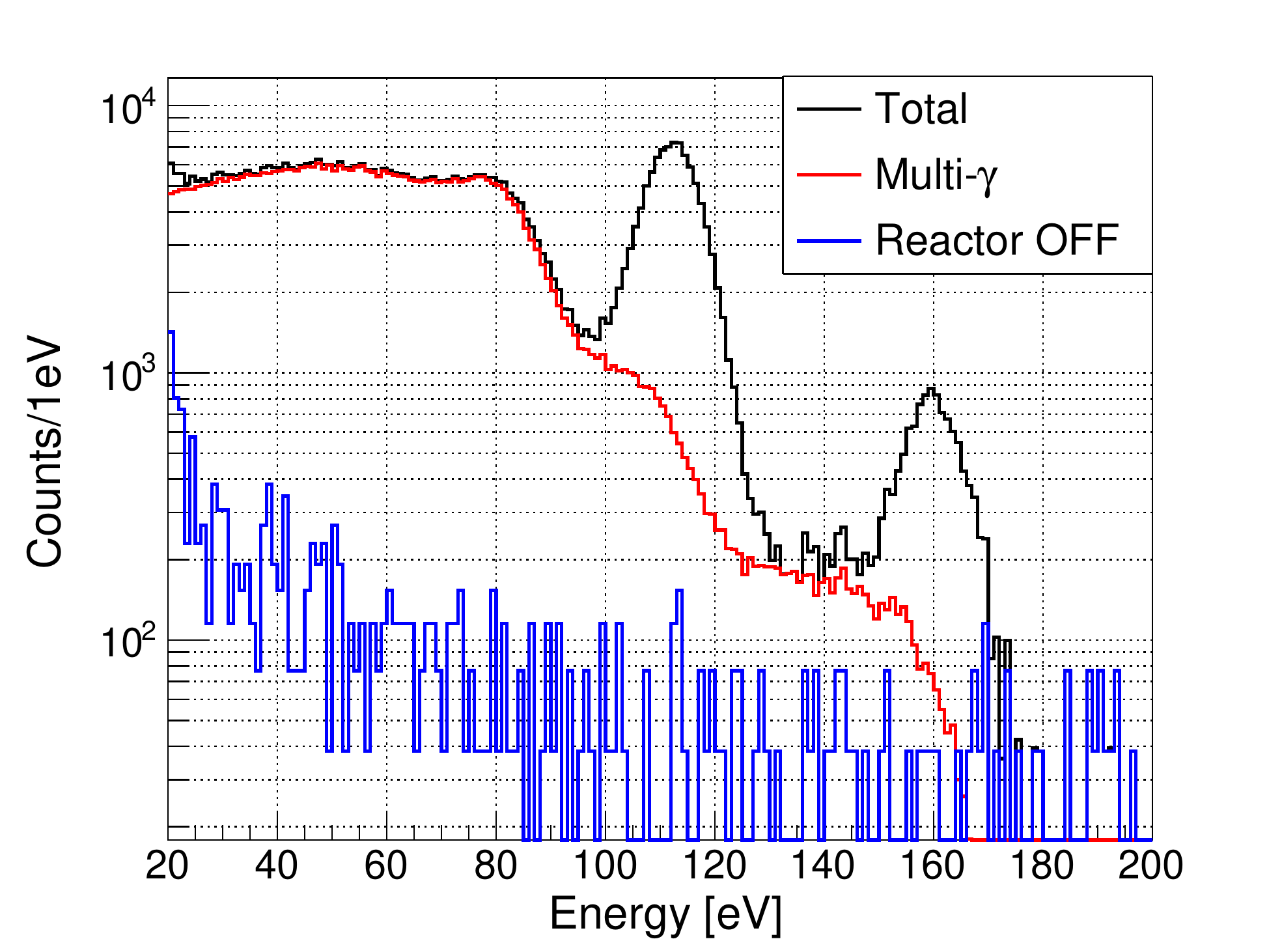}
	\caption{\label{fig:CaWO4_Signal}Simulated response of a 0.76 g CaWO$_4$ (5\,mm length cube) detector with 5 eV energy resolution, exposed to a thermal neutron flux of 270 n/cm$^2$/s during 3.4 days. The red and blue spectra correspond to the main background contributions in the RoI, labelled 1) and 3) in table \ref{tab:rates} respectively. The “Reactor OFF” data, blue points, are taken from \cite{Strauss:2017cam} and re-scaled to the simulated acquisition time. One notes that in the total spectrum (black) the calibration peaks at 112 eV and 160 eV, associated to the single-$\gamma$ transition of $^{183}$W and $^{184}$W respectively, are clearly visible above background.}
	\end{center}
\end{figure}

 The expected spectra of energy deposits in the RoI are shown in figure~\ref{fig:CaWO4_Signal} for a  run time of 3.4 days, with a \NUCLEUS CaWO$_4$ crystal. The high energy resolution of the detector, taken here as 5 eV, is critical in keeping the mono-energetic calibration events in a small energy band, helping to suppress the more spread-out backgrounds. The peaks corresponding to the single-$\gamma$ transitions of $^{183}$W and $^{184}$W are clearly visible at 112.5~eV and 160.3~eV respectively, with a signal-to-background ratio around 10. This makes this calibration method very robust for the CaWO$_4$ case. With a 5 eV energy resolution ($\sigma$), the mean position of this peaks could be determined with 1\% statistical accuracy in less than one hour of data taking. 
 
 According to figure \ref{fig:Edep_Bolo_W} a third calibration peak is expected at 85.8~eV from the $^{187}$W isotope. However in figure \ref{fig:CaWO4_Signal} the contribution of the single-$\gamma$ transitions is overwhelmed by the large step of multi-$\gamma$ events seen in the spectrum around the same energy. We will discuss in the next section how the other experimental variant, namely detection of a high-energy $\gamma$ in coincidence with the signal in the cryodetector, can be used to recover a calibration peak from this configuration and in the same time is mandatory to extend the application of this method to germanium bolometers.
 
 At this point we should stress that the prediction of the recoil spectra are rather sensitive to the inputs from the nuclear data bases, which feed into the nuclear level simulations. Here we have used the EGAF database because it includes the most recent evaluations of the transitions of interest and especially the measured $S_n$ to ground state transitions \cite{Hurst2014}. If we use instead other databases such as ENSDF $S_n$ to ground state transitions are missing for some isotopes and then their branching ratios only rely on models. We observed that in the ENSDF database the 160 eV peaks is weaker and the 87 eV peak stronger while the main 112 eV peak is about constant and always clearly present. Similar variations can apply for the germanium results below and for other isotopes. Note however that this uncertainty in the nuclear data concerns only the branching ratio of the transitions while their energies are well known and in very good agreement from one database to another. Thus, as long as the signal-to-noise ratio remains favorable for the extraction of a calibration peak, the determination of its position remains accurate.  The measurement of relative amplitudes between peaks can then be used as a strong constraint on branching ratios.

\subsection{``Photon tagging''-- Single-$\gamma$  coincident with cryogenic detector}
\label{subsec:coinc-mode}

For the simulation, two cylindrical BGO crystals ($\oslash 3^{''}\times3^{''}$) are placed on both sides of the detector at 4 cm distance from its surface (figure~\ref{fig:Exp_Concept}). These $\gamma$-detectors can be operated in a cryostat and provide a good compromise between rate, energy resolution, detection efficiency, compactness, and timing of the light signal (detected by a bolometric photodetector). Their  performance is taken from measurements with a $\oslash$20$\times$20 mm$^3$ BGO detector \cite{Marcillac:2003nat} operated at 20 mK. In particular a 2.2\% energy resolution ($\sigma$) was obtained for the $^{208}$Tl line at 2.615 MeV, from which we extrapolate the resolution at other energies assuming a $\sqrt{E}$ law. 

\begin{figure}[ht]
    \begin{center}
	\includegraphics[width=0.8\linewidth]{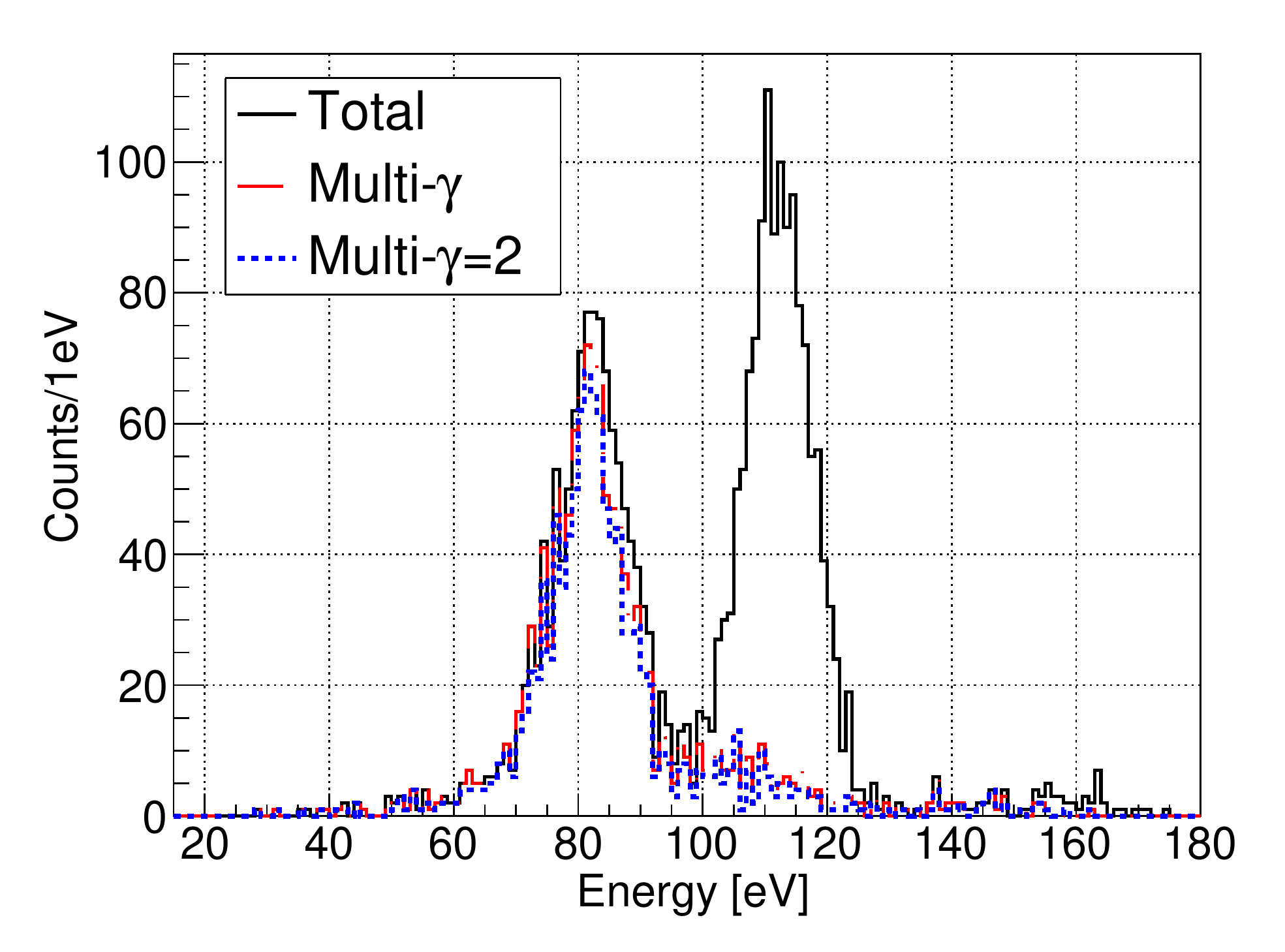}
	\end{center}
	\caption{Effect of photon tagging. Black solid line: with a \NUCLEUS CaWO$_4$ bolometer, an energy of 5.47 $\pm$ 0.2 MeV, corresponding to the transition from $S_n$ of $^{187}$W, is required in one of the BGO detectors. The associated peak becomes clearly visible and centered at 82.4 eV, while the peak on the right corresponds to the $^{183}$W transition shown in figure \ref{fig:CaWO4_Signal} where its single-photon (6.19 MeV) is registered in the 5.47 $\pm$ 0.2 MeV window. Dashed-red and dotted-blue lines: recoil distributions induced by the emission of more-than-one and two $\gamma$'s respectively. The 2-$\gamma$ contribution is dominant in the 80 eV region and at the origin of the slight shift of the mean value with respect to 85.6 eV, expected from single-$\gamma$ transitions (see text for details).}
	\label{fig:Coinc_CaWO4}
\end{figure}

We then request a coincident $\gamma$ in one of the  BGO crystals with energy 5.467~$\pm$~0.2 MeV. The central value is the energy of the single-$\gamma$ transition of $^{187}$W, whose associated nuclear recoil at 85.6 eV is smeared out in the bolometer-only spectrum of figure \ref{fig:CaWO4_Signal}. The width of $\pm$0.2~MeV corresponds to about $\pm$2 $\sigma$ of the energy resolution of the BGO. Figure \ref{fig:Coinc_CaWO4} shows how the coincidence technique makes a new calibration peak clearly visible in the 80 eV region. However the mean value, 82.4 eV, is a few eV below the expected 85.6 eV. This is explained by the fact that most of the recoils populating this peak are induced by the emission of 2 $\gamma$-rays (dotted blue curve in figure \ref{fig:Coinc_CaWO4}). Indeed, the EGAF database indicates that the branching ratio towards single-$\gamma$ transitions is very low for $^{187}$W, 0.26\% from table \ref{tab:nuc_data}. Instead, transitions with a specific configuration of 2 $\gamma$'s are quite probable: one $\gamma$ is emitted with an energy close to $S_n$ and the second $\gamma$ carries the small missing energy (about 150 keV) to reach the ground state. The resulting distribution of recoils spans a range delimited on the high-recoil edge by the same emission direction of the two $\gamma$'s (85.6 eV recoil) and on the low-recoil edge by the opposite emission directions (about 75 eV recoil). A selection window of the BGO events centered on the full $S_n$ energy favors the “close-to-same emission direction” configurations where both $\gamma$'s reach the detector therefore the shift of the mean position is small, about 3 eV here. On the contrary, moving the central position of the selection window of the $\gamma$ energy to few 100 keV downward disfavors the “same emission direction” configuration and reduces further the mean position of the peak by about 2 eV. This can be used in the analysis to pinpoint the process at work and thus ensure a measurement of this third calibration peak with an accuracy similar to the other peaks. Such measurement would extend the CaWO$_4$ calibration to lower energy and provide a unique opportunity to test the local linearity of the phonon signal in the unexplored range of 100 eV.

Of course the same photon-tagging method can be used for the two other peaks at 112 eV and 160 eV. These peaks were already clearly visible in the bolometer spectrum (figure~\ref{fig:CaWO4_Signal}). Requesting a coincidence with the high-energy $\gamma$, it improves further the signal-to-background ratio by a factor 3 and makes the small residual background contribution almost symmetrical around the mean peak position. Such configuration is then suitable for high precision measurement of the peak position even without any knowledge of the $\gamma$-cascades.

\begin{figure}[ht]
    \begin{center}
	\includegraphics[width=1\linewidth]{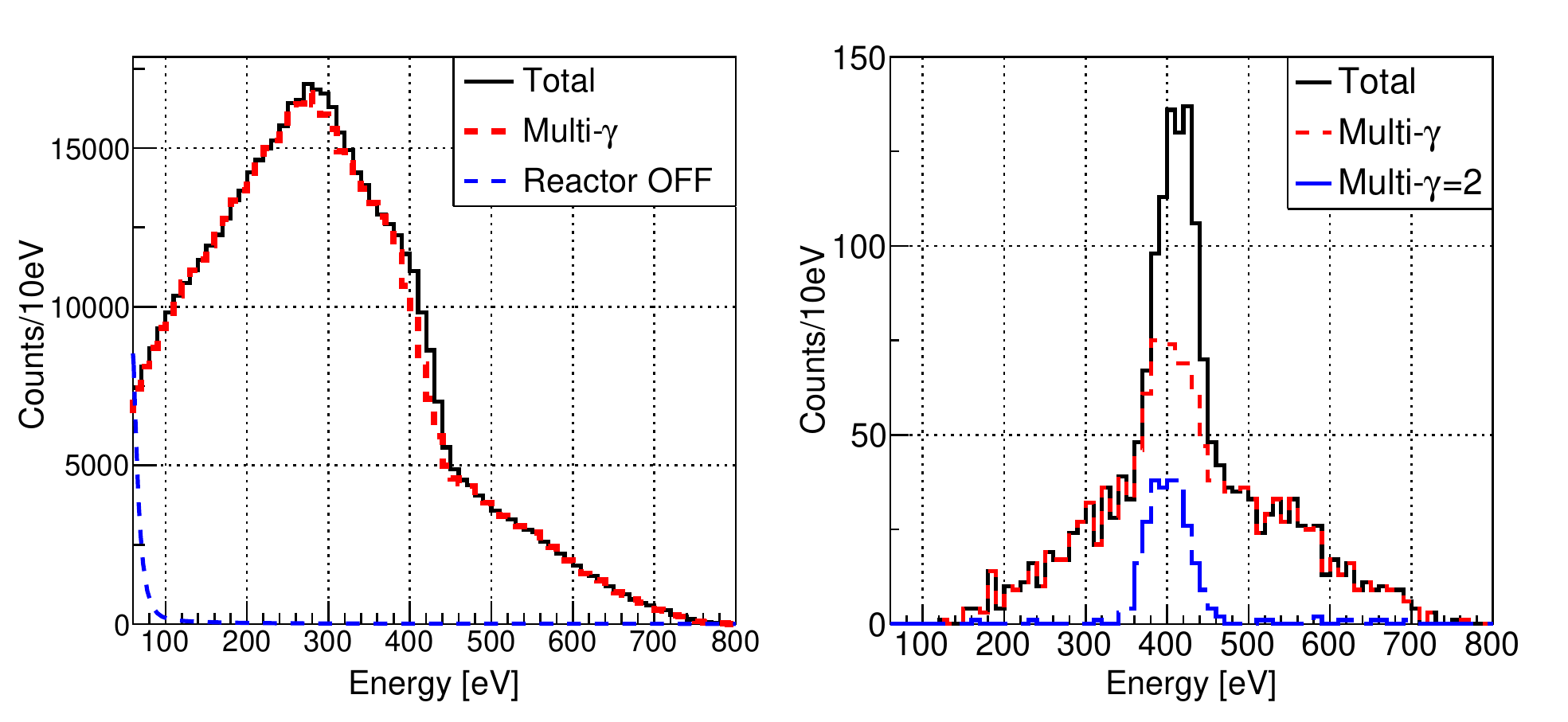}
	\end{center}
	\caption{Photon tagging with a Ge bolometer. Simulation of a 33 g Ge low-temperature detector in an incident flux of 5 n/cm$^2$/s (corresponding to about 2.7 neutron-capture/s) for 7 days and a 20 eV energy resolution (1 $\sigma$). Left: Cryodetector spectrum. The “Reactor OFF” background is assumed as measured in \cite{Armengaud:2019kfj}. Right: Same simulation when an energy in the [7.2--7.6] MeV range is requested in one of the two $\gamma$-detectors. }
	\label{figs:Ge_Single_Coinc}
\end{figure}

Germanium is also of particular interest due to its widespread use for the detection of low-energy recoils ~\cite{Hakenmuller:2019ecb,Agnese:2017jvy,Armengaud:2019kfj,Aalseth:2012if,Arnaud:2020svb}. We have simulated the recoil spectrum in a Ge detector, considering the main capture isotopes listed in table~\ref{tab:nuc_data} and assuming energy ($\sigma$=20 eV resolution), time responses and “reactor OFF” background as measured by the \EDELWEISS collaboration \cite{Armengaud:2019kfj}. The $^{74}$Ge calibration peak, expected at 754 eV, is unfortunately suppressed because of the high multipolarity of the $S_n$ to ground-state transition. The main calibration signal, expected at 416 eV from the single-$\gamma$ transitions of $^{71}$Ge is visible in the recoil spectrum but largely dominated by the multi-$\gamma$ background. Figure \ref{figs:Ge_Single_Coinc} illustrates how a coincidence with a high energy $\gamma$-ray recovers the 416 eV calibration peak. The cut for the energy in the BGO detectors is centered on 7.416 MeV, the nominal energy of the single-$\gamma$ transition of $^{71}$Ge, with the same $\pm$0.2 MeV width. The simulations clearly show the three main selected components: the sought-for single-$\gamma$ signal, a two-$\gamma$ background with one high-energy $\gamma$ close to $S_n$ as for $^{187}$W and a flatter background from events with more than two $\gamma$ emitted. As for the CaWO$_4$ case discussed above, we note that this coincidence technique not only suppresses the multi-$\gamma$ background but also makes its distribution a lot more symmetrical around the position of the signal peak such that only a modest knowledge of the residual background is needed for calibration. For instance, a simple fit with two Gaussians, one for the calibration peak (assuming that the single and two-$\gamma$ components are centered on the same value) and the other for the smeared-out multi-$\gamma$ background, is enough to recover the correct calibration energy at the 1\% level.

\section{Conclusion and perspectives}
\label{sec:conclusion}
We present an experimental concept for the calibration of low-temperature particle detectors in the 100\,eV range of nuclear recoils, suitable for experiments to search for low-mass dark matter particles and/or coherent neutrino-nucleus scattering. The measurement is based on the irradiation of the detector by a thermal neutron flux. After capture of a neutron, a target nucleus has a non-negligible probability to de-excite by the emission of a single $\gamma$-ray, thus producing a calibration line of nuclear recoils. We have demonstrated the unique feature of this calibration method to combine an accuracy on the scale of one percent and a uniform exploration of the bolometer volume with nuclear recoils in the 100 eV range. 

Here we have studied the situation for complex nuclei, as relevant to a number of projects using W-based cryogenic detectors, such as \NUCLEUS$\!$ and \CRESST~\cite{Abdelhameed:2019hmk} or Ge cryogenic detectors, such as \EDELWEISS~\cite{Armengaud:2019kfj}. The case of tungsten was found very favorable with strong calibration lines above background expected in the raw spectrum of a \NUCLEUS bolometer (5 mm length cube). The nuclear properties of the Ge nucleus are less favorable and the volume of the detector considered here is about 40 times larger leading to a 4 times larger energy resolution (20 eV instead of 5 eV). In that case no peak is visible in the raw detector spectrum but we show that requesting a coincidence with the emitted $\gamma$ in a BGO crystal allows the accurate measurement of a calibration peak at about 400 eV within few days of measurement.

This work suggests a number of interesting issues for further study. We will examine other detector materials, the situation for the lighter nuclei is studied in a parallel paper~\cite{leo2020}. Beyond the simplest case of a single emitted $\gamma$-ray defining a recoil line, the detection of events associated with two-$\gamma$ cascades will be investigated. These events can be selected requesting two narrow energy windows in two distinct $\gamma$ detectors, with the sum of these two energies centered on the $S_n$ value. From the conservation of momentum the mean value of the selected nuclear recoils becomes a function of the angle between the two $\gamma$ detectors, with a minimal value reached in the “back-to-back” configuration. Such triple coincidences between the central detector and two $\gamma$-rays would offer an extended calibration range toward lower recoil energies, at the price of a strongly reduced counting rate to be mitigated by an efficient and large acceptance $\gamma$ detection.

The calibration method presented will find further applications. The neutron capture technique could considerably reduce the uncertainties in the low-energy ionization quenching factor of Ge detectors \cite{BARKER20121} currently limiting their sensitivity at low energies ~\cite{Hakenmuller:2019ecb}. Accessing the  nuclear recoil line at around 400\,eV (see Fig. \ref{figs:Ge_Single_Coinc})  with a benchmark Ge cryogenic detector, which measures simultaneously charge and phonons ~\cite{Agnese:2018gze}, would enable an ionization quenching measurement at sub-keV energies for the first time. 
Furthermore, to disentangle possible phonon quenching effects at low energies the calibration of nuclear recoils will be combined with state-of-the-art techniques for electron-recoil calibration in the sub-keV energy range, such as X-ray fluorescence sources and absolute energy calibration with LED pulses \cite{Cardani:2018krv}. Simultaneous measurements with the same cryogenic detector will allow to establish direct correlations between energy scales for nuclear and electron recoils. This will enable to disentangle the relevant atomic and nuclear physics effects at sub-keV energies. While above a few keV, the detector response of cryogenic detectors is well understood and independent of the type of particle interaction~\cite{Strauss:2017cuu,Abdelhameed:2019hmk}, quenching effects might be of high relevance for the signal reconstruction at sub-keV energies. That's in particular relevant, when approaching the lattice dislocation energies of detector crystals at typically $\mathcal{O}(10\ \text{eV})$.  
 Recoils in this regime are expected to fundamentally change the phonon processes. The struck nuclei can remain in their lattice positions followed by purely phononic excitations, while more energetic nuclear recoils  induce various additional processes, such as ionization or lattice damage. 
Finally, this technique could also be used to study and validate directional sensitivity in cryogenic detectors~\cite{PhysRevD.98.115034} which would allow to search for a diurnal modulation of dark matter signal and to disentangle reactor or solar neutrinos from limiting backgrounds. Defect creation and the effective displacement energies depend on the recoil angle with respect to the crystal-lattice orientation~\cite{PhysRevLett.120.111301}. The detection of high energy $\gamma$-rays in coincidence with the bolometer signal should give sensitivity to the direction of the initial nuclear recoil and provide a unique calibration method for future directional low-threshold detectors which might be crucial for next-generation rare-event searches.

\bibliographystyle{unsrtnat}
\bibliography{./crab_paper}

\end{document}